\documentclass[11pt]{article}

\usepackage[english]{babel}
\usepackage[utf8]{inputenc}
\usepackage[T1]{fontenc}
\usepackage[a4paper,top=1.75cm,bottom=1.8cm,left=2.1cm,right=2.1cm]{geometry}
\usepackage{setspace}
\usepackage{amsmath}
\usepackage{graphicx}
\usepackage[colorlinks=true, allcolors=blue]{hyperref}
\usepackage{enumitem}
\usepackage{booktabs}
\usepackage{microtype}
\usepackage{titlesec}
\usepackage[font=small,labelfont=bf]{caption}

\titlespacing*{\section}{0pt}{0.7em}{0.35em}
\titlespacing*{\subsection}{0pt}{0.55em}{0.25em}
\setlist[itemize]{leftmargin=1.2em,itemsep=0.1em,topsep=0.2em,parsep=0pt}
\setlist[enumerate]{leftmargin=1.4em,itemsep=0.1em,topsep=0.2em,parsep=0pt}
\usepackage{etoolbox}
\makeatletter
\patchcmd{\thebibliography}{\list}{\footnotesize\list}{}{}
\patchcmd{\thebibliography}{\setlength{\itemsep}{\z@}}{\setlength{\itemsep}{0pt}\setlength{\parsep}{0pt}}{}{}
\makeatother

\newcommand{\hst}{\emph{Hubble Space Telescope}}
\newcommand{\hwo}{\emph{Habitable Worlds Observatory}}
\newcommand{\gaia}{\emph{Gaia}}
\newcommand{\romanst}{\emph{Roman}}

\title{\vspace{-1.0em}\bfseries From Images to Orbits:\ A Hubble 2030s Astrometric Legacy for the M31--M33 Star Cluster Systems}
\author{Borja Anguiano$^{1}$, Benjamin J.\ Gibson$^{2}$, Sten Hasselquist$^{2}$\\[-0.1em]
{\small $^{1}$Centro de Estudios de F\'isica del Cosmos de Arag\'on (CEFCA), Plaza San Juan 1, 44001 Teruel, Spain}\\
{\small $^{2}$Space Telescope Science Institute (STScI), 3700 San Martin Dr, Baltimore, MD 21218, USA}\\
[-0.1em]
{\small Contact: \texttt{banguiano@cefca.es}}}
\date{White paper submitted to the STScI call for Hubble science into the 2030s \\ May 2026}

\begin{document}
\maketitle
\vspace{-1.2em}

\begin{abstract}
The most irreplaceable capability of \hst\ in the 2030s is not only its angular resolution or its UV--optical sensitivity, but its accumulated time baseline. We recommend a Hubble Local Group Astrometric Legacy that would obtain matched 2030s ACS/WFC and WFC3/UVIS imaging of selected M31 and M33 star cluster fields, combine those observations with a careful astrometric audit of the existing archive, and deliver public transverse velocity products through MAST. PHAT and PHATTER already provide exceptional first epochs across the nearest large external spirals, resolving tens to hundreds of millions of stars and identifying thousands of star clusters and compact background reference objects. Earlier targeted HST programs, dating back to the mid-1990s, can extend the temporal baseline in selected cases, but only after camera, chip, filter, dithering, crowding, and reference-frame triage. A coherent 2030s repeat campaign would turn the best of this archive into bulk cluster motions, enabling the first systematic orbital mapping of star cluster populations beyond the Milky Way. These measurements would constrain disk heating, cluster disruption, accretion histories, stream associations, the M31--M33 interaction, and the mass distributions of Local Group spirals. The same program would also preserve and stress-test the calibration infrastructure needed for high-stability optical/UV astrophysics in the \hwo\ era: geometric-distortion solutions, PSF and CTE modeling, matched-epoch observing strategies, long-term reference frames, and durable public high-level products.
\end{abstract}

\section{Executive recommendation}

We recommend a \textbf{Hubble Local Group Astrometric Legacy} initiative: a coordinated program of repeat, astrometrically optimized imaging of selected M31 and M33 star cluster fields with ACS/WFC and WFC3/UVIS during Hubble's extended lifetime into the 2030s.  The program should be designed from the start to produce public high-level science products: cluster-frame proper motions, local distortion and PSF solutions, membership probabilities, cross-identifications with PHAT/PHATTER catalogs, and reference-frame products tied to compact background galaxies and, where useful, \gaia\ foreground stars.

PHAT mapped roughly one third of M31's star forming disk in six HST filters, resolving more than $10^8$ stars over a contiguous $\sim0.5~\mathrm{deg}^2$ area \cite{Dalcanton2012}; the PHAT cluster search identified 2,753 clusters and 2,270 background galaxies in the searched footprint \cite{Johnson2015}.  PHATTER extended the same UV--optical--near-IR strategy to M33, delivering photometry for $\sim22$ million stars \cite{Williams2021} and a catalog of 1,214 star clusters \cite{Johnson2022}.  These data are a first epoch of exceptional quality.  They are also not the only first epoch: the HST archive contains targeted cluster observations from earlier programs, including WFPC2 M31 globular cluster imaging \cite{Rich1996,Rich2005}, archive-based M31 cluster surveys and ACS/RBC cluster photometry \cite{Barmby2001,Krienke2008,Perina2009}, deep ACS pointings such as SKHB-312 \cite{Brown2004}, and M33 WFPC2/ACS cluster fields and catalogs from other archival surveys \cite{Sarajedini1998,Sarajedini2000,Sarajedini2007,SarajediniMancone2007,ParkLee2007,SanRoman2009}.  If Hubble survives and is scientifically prioritized into the 2030s, these heterogeneous first epochs become a multi-decade astrometric experiment, provided the roadmap separates precision-grade ACS/WFC and WFC3/UVIS material from WFPC2 fields that require case-by-case calibration and quality control.

Not all first epochs should be weighted equally.  ACS/WFC and WFC3/UVIS fields are the natural precision anchors for new 2030s repeats; ACS/HRC can be useful where it supplies an archival first epoch, but its field is small; WFPC2 is scientifically valuable because of its long baseline but is not a drop-in equivalent to later cameras.  WFPC2 fields should be promoted to the legacy sample only if they have adequate sampling, dithering, chip placement, exposure depth, and inertial or hybrid reference objects.  This triage is a strength of the proposal: it makes the program realistic rather than overclaiming the full archive.

This white paper addresses the four questions in the STScI call \cite{STScI2030}.  The science requires Hubble because no other current or planned facility can reproduce Hubble's own multi-decade, diffraction-limited UV--optical baseline in crowded Local Group fields.  The needed capabilities are stable ACS/WFC and WFC3/UVIS imaging, careful calibration, matched observing modes, and archive-supported astrometric pipelines.  The HWO connection is direct: this program would translate a flagship science use case---stable, high-resolution UV--optical imaging of crowded fields over decade baselines---into calibration, operations, and archive requirements for a future ``super Hubble.''

\section{The science challenge: cluster systems as orbit tracers}

The star clusters of M31 and M33 are compact, luminous, age-datable, and chemically interpretable tracers of galaxy assembly. Clusters carry population-level information: their ages, metallicities, locations, morphologies, and orbits jointly encode whether they formed in disks, were heated by interactions, or were accreted with satellites.  Their transverse motions are therefore a direct route from resolved Hubble images to galaxy-scale dynamics.

A Hubble 2030s astrometric legacy program would address several linked questions:

\begin{itemize}
\item \textbf{How do cluster orbits separate in-situ formation from accretion?}  Old clusters associated with stellar streams or halo substructures should exhibit orbital coherence that distinguishes accreted populations from long-lived disk or bulge populations.
\item \textbf{How are young clusters dynamically processed?}  Motions of young and intermediate-age clusters in M31 and M33 would test disk heating, spiral-arm perturbations, cluster disruption, and the dynamical role of rings, bars, and warps.
\item \textbf{What was the dynamical history of the M31--M33 interaction?}  Competing scenarios for the recent interaction history of M31 and M33 predict different patterns of disk disturbance, halo substructure, and cluster orbital families \cite{Patel2025}.
\item \textbf{How massive are the halos of M31 and M33?}  Tangential velocities of clusters and satellites complement line-of-sight velocities and reduce degeneracies in mass modeling, especially when combined with streams and satellite proper motions \cite{vanDerMarel2012,Patel2023}.
\item \textbf{Can Hubble create the first external-galaxy analog of Galactic cluster orbital archaeology?}  The Milky Way now uses 6D phase-space information to connect clusters to accretion events.  Hubble can extend that logic to the nearest large spirals.
\end{itemize}

At the distances of M31 and M33, $10~\mu\mathrm{as}~\mathrm{yr}^{-1}$ corresponds to $\sim37$--$40~\mathrm{km~s}^{-1}$, using $v_t = 4.74\,\mu D$.  This sets the appropriate ambition.  The primary goal should be \emph{bulk motions of cluster populations and cluster-associated stellar samples}, not internal velocity dispersions for most individual clusters.  Internal motions may be possible for exceptional nearby, massive, or well-resolved cases, but they should be a secondary objective.  \textbf{The robust legacy product is a catalog of cluster-system transverse motions, with uncertainties to distinguish disk rotation, halo orbits, stream membership, and interaction-driven kinematic families.}

\section{Why Hubble is required in the 2030s}

Hubble proper motion measurements in the Local Group have already demonstrated the essential method: compare high-S/N stellar positions across epochs, measure motion relative to compact background galaxies and/or foreground stars with known motions, and average over many stars to control random errors.  HST measured the proper motion of M31 using 5--7 year baselines, thousands of M31 stars, and hundreds of compact background galaxies per field, reaching a final accuracy of $0.012~\mathrm{mas~yr}^{-1}$ \cite{Sohn2012}.  Multi-epoch HST imaging has also measured the proper motions of M31 satellites such as NGC 147 and NGC 185, with orbital implications for their connection to M31 halo substructure \cite{Sohn2020}; recent HST+Gaia approaches show how Gaia foreground stars and compact background galaxies can be combined in practical reference frame solutions \cite{Warfield2023}.

\begin{figure}[t]
\centering
\includegraphics[width=\linewidth]{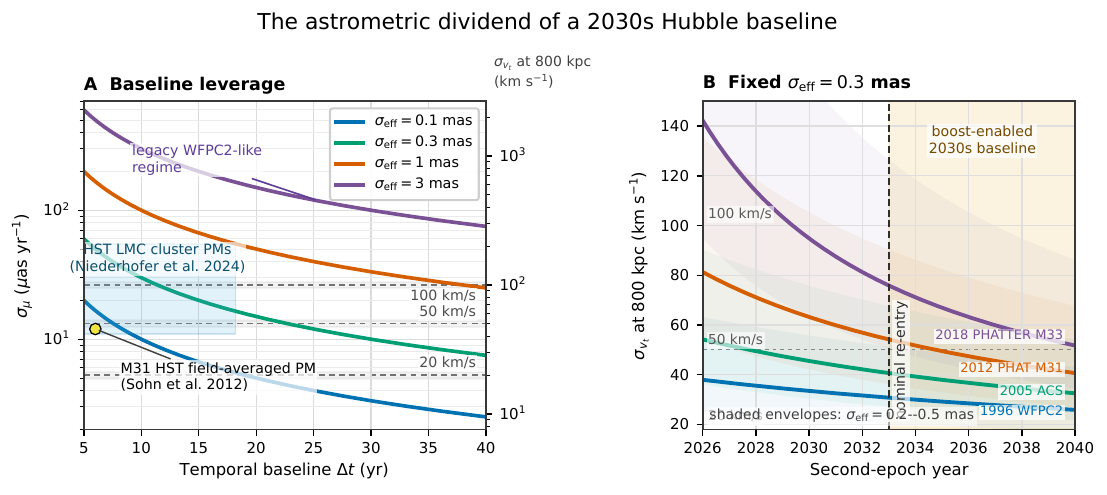}
\vspace{-0.8em}
\caption{\label{fig:astrometric_dividend}\textbf{The astrometric dividend of a 2030s Hubble baseline.} Left: proper motion precision improves linearly with temporal baseline; the equivalent transverse velocity scale at $D=800$ kpc is shown for representative effective two-epoch displacement uncertainties, $\sigma_{\rm eff}$. Literature benchmarks show that HST proper motions already reach $\sim$10--50 $\mu\mathrm{as}~\mathrm{yr}^{-1}$ in favorable multi-epoch ACS/WFC and WFC3/UVIS data \cite{Sohn2012,Niederhofer2024}. Right: for fixed $\sigma_{\rm eff}=0.3$ mas, a second epoch in the late 2030s improves existing M31/M33 archival cohorts from marginal or coarse transverse velocities toward the $\sim50~\mathrm{km~s}^{-1}$ bulk-orbit regime. WFPC2-era targeted clusters provide the longest baselines, but only after astrometric-quality triage; PHAT-era M31 and PHATTER-era M33 fields gain substantial leverage if Hubble operates beyond the nominal 2033 re-entry date.}
\vspace{-0.9em}
\end{figure}

Figure~\ref{fig:astrometric_dividend} illustrates the central payoff: a Hubble fifth decade does not merely add another observing epoch; it changes the transverse velocity precision accessible from existing M31/M33 archival imaging.

\newpage

The key limitation is not proof of concept, but baseline and program design.  Hubble images obtained in the 1990s, 2000s, and 2010s become dramatically more powerful when repeated in the 2030s: a 20--35 year baseline reduces the required per-epoch positional precision and helps separate real motions from residual systematics.  The relevant first epochs are heterogeneous rather than uniform: Rich-era M31 globular cluster pointings, Barmby/Krienke/Perina archival cluster fields, Brown's deep ACS SKHB-312 field, Sarajedini M33 halo-cluster images, Park/Lee WFPC2 archive fields, and San Roman radial ACS fields are candidates for a MAST audit and prioritized repeats.  But heterogeneity is not a virtue by itself.  ACS/WFC, ACS/HRC, and WFC3/UVIS have been the workhorses for high-precision HST proper motion catalogs \cite{Bellini2014}; the M31 velocity-vector work used ACS/WFC and WFC3/UVIS with explicit CTE, PSF, and distortion treatment, including ACS/WFC corrections at the $\sim0.01$ pixel level and WFC3/UVIS corrections better than $0.008$ pixel \cite{Sohn2012,Bellini2011}.  WFPC2 first epochs are valuable mainly because of their age, but should be used only after instrument-specific triage: modern recalibration finds low-frequency WFPC2 residuals of order $1.4$--$2.3$ mas on the PC chip and $2$--$3$ mas on the WF chips \cite{CasettiDinescu2021}.  Thus WFPC2 is a baseline extender, not the default precision camera.

Ground-based facilities, even with adaptive optics, cannot reproduce the stable, wide-field, optical/near-UV Hubble archive in these crowded fields.  JWST provides extraordinary infrared sensitivity, but not a direct replacement for Hubble's optical/UV first epochs, detector geometry, or matched ACS/WFC and WFC3/UVIS astrometric legacy.  \romanst\ will provide complementary wide-field near-IR imaging and could extend the reference-frame context, but Hubble is needed to connect the existing UV--optical images to the 2030s.

\section{Required capabilities and operational priorities}

The proposed program is not a request for new hardware.  It is a request to preserve and prioritize the specific Hubble capabilities that make multi-decade astrometry possible.

\begin{table}[h]
\centering
\caption{Required Hubble capabilities for a M31--M33 cluster astrometric legacy program.}
\small
\begin{tabular}{p{0.28\linewidth}p{0.64\linewidth}}
\toprule
\textbf{Requirement} & \textbf{Why it matters for M31--M33 cluster astrometry} \\
\midrule
ACS/WFC and WFC3/UVIS repeat imaging & These are the default 2030s cameras: they best match PHAT/PHATTER optical and near-UV epochs and minimize color-, PSF-, and distortion-dependent systematics. \\
Archival instrument triage & WFPC2, especially WF-chip imaging, should be used as a long-baseline first epoch only after checks on sampling, dithering, saturation, crowding, background galaxies, filter match, and detector placement. \\
Stable geometric-distortion solutions & Residual distortion is a dominant systematic for microarcsecond-per-year proper motions; time-dependent and filter-dependent solutions must remain supported, including WFC3/UVIS realignment to Gaia where possible \cite{OConnor2024}. \\
PSF and CTE calibration & Cluster fields are crowded and heterogeneous; spatially variable PSFs and charge-transfer effects can bias centroids if not modeled consistently across epochs. \\
Matched observing strategy & Repeat imaging should use matched filters, pointings, roll angles, dithers, exposure times, and detector placement where feasible, while avoiding saturation of useful reference stars. \\
Reference-frame design and data products & Fields need compact background galaxies and unsaturated Gaia stars; MAST products should include local transformations, proper motion catalogs, quality flags, systematic-error floors, and reproducible notebooks. \\
\bottomrule
\end{tabular}
\end{table}

Operationally, the highest priority is to treat these observations as an astrometric experiment, not simply as repeat imaging.  The camera hierarchy should be explicit: (1) ACS/WFC and WFC3/UVIS define the preferred second epoch; (2) ACS/WFC, ACS/HRC, and WFC3/UVIS first epochs are highest-priority repeats; (3) WFPC2 first epochs, including Rich/Sarajedini-era cluster fields, are high-value but opportunistic; and (4) WFC3/IR, NICMOS, and STIS are special-case/context data rather than the core PM basis.  Target fields should maximize archival leverage, cluster density, background-galaxy availability, filter match, and science diversity.

\section{A major initiative: the Hubble Local Group Astrometric Legacy}

A scientifically powerful implementation would be a multi-cycle Treasury- or Legacy-scale program with four tiers:

\begin{enumerate}
\item \textbf{Core matched-epoch fields.}  Reobserve a carefully chosen subset of PHAT and PHATTER cluster-rich fields with the closest practical match to the first-epoch imaging strategy.  These are the precision astrometry anchors.
\item \textbf{Archival cluster-baseline fields.}  Reobserve selected M31/M33 clusters with calibration-qualified older ACS or WFPC2 epochs, including Rich-era M31 globular clusters, Barmby/Krienke/Perina fields, deep Brown ACS fields, Sarajedini M33 halo clusters, Park/Lee WFPC2 archive fields, and San Roman ACS radial fields.  These are not uniform survey fields; their value depends on both baseline and astrometric quality.
\item \textbf{Stream and halo fields.}  Add fields containing old clusters, halo substructure, and compact background galaxies to connect cluster motions to the accretion history of M31 and M33.
\item \textbf{Calibration and transfer fields.}  Include fields optimized for Gaia-star reference-frame transfer, background galaxy centroiding, filter-dependent distortion tests, and cross-instrument calibration with \romanst\ and JWST where appropriate.
\end{enumerate}

The community deliverables should be explicitly defined before observing begins.  At minimum they should include: (i) public proper motion catalogs for clusters and cluster associated resolved stars; (ii) local astrometric transformations, covariance information, and camera-specific systematic error floors; (iii) machine-readable crossmatches to PHAT, PHATTER, Gaia, spectroscopic, and variable-star catalogs; (iv) calibrated image stacks and masks for saturated stars, diffraction spikes, blends, and background galaxies; and (v) a reproducible pipeline documented well enough for use by the broader Local Group community.  Hubble has already shown through PHAT, PHATTER, OPAL, ULLYSES, and other legacy programs that high-level products multiply the value of the original observations.  A 2030s astrometric legacy should be designed with the same philosophy.

The program would also be synergistic with facilities that will shape the next decade.  \romanst\ can supply wide-field near-IR context, variable-star and stellar-population maps, and a complementary astrometric grid.  JWST can provide deep infrared follow-up for the most crowded or extinguished fields.  Ground-based spectroscopy from Keck, Subaru/PFS, WEAVE, DESI, SDSS-V and future extremely large telescopes can provide radial velocities and abundances.  Hubble supplies the transverse component that turns those data into orbits.

\section{Preparing for HWO}

A Hubble 2030s Local Group astrometric legacy would be an end-to-end pathfinder for the broader astrophysics portfolio of a future UV--optical--IR flagship.  NASA describes HWO as a large infrared/optical/ultraviolet space telescope that will search for life-bearing worlds and also provide a platform for transformational astrophysics \cite{NASA_HWO}.  The same infrastructure that will make HWO a powerful ``super Hubble'' for astrophysics---stable image sampling, calibrated detector geometry, long-term reference frames, PSF control, and durable archives---is exactly what multi-decade crowded-field astrometry requires.

Hubble can prepare HWO in four practical ways.  First, it can quantify the calibration accuracy needed for microarcsecond-per-year astrometry in real crowded fields.  Second, it can define observing modes that preserve both point source and compact galaxy reference frames.  Third, it can demonstrate how high-level astrometric products should be archived so future users can combine data separated by decades.  Fourth, it can provide a science driven benchmark for HWO trades involving field of view, sampling, stability, wavelength coverage, detector systematics, and pipeline design.  In this sense, a Local Group astrometric legacy is a concrete test of the calibration and archive practices that HWO's broad astrophysics program will require.

\section{Concluding statement}

The proposed Hubble 2030s roadmap should recognize time baseline as a unique facility capability, while also recognizing that not every archival image is a valid astrometric first epoch.  For M31 and M33, the best first epochs already exist, from PHAT/PHATTER and selected older targeted cluster programs; the stars and clusters are already resolved; the background reference objects are already embedded in many of the images.  What is missing is a coherent 2030s strategy to repeat the right fields in the right modes and to deliver the resulting motions as a public legacy.  A Hubble Local Group Astrometric Legacy would transform PHAT and PHATTER from imaging surveys into orbital surveys, provide the first systematic transverse velocity map of external galaxy cluster systems, and establish calibration standards directly relevant to HWO.  This is a compelling, cost-effective use of Hubble's possible fifth decade.

\end{document}